\documentclass{elsart}

\usepackage{epsfig}

\usepackage{bm} % bold math
\usepackage{mathptmx} % postscript fonts
\newcommand{\be}{\begin{equation}}
\newcommand{\ee}{\end{equation}}
\newcommand{\bc}{\begin{center}}
\newcommand{\ec}{\end{center}}
\newcommand{\bi}{\begin{itemize}}
\newcommand{\ei}{\end{itemize}}
\newcommand{\ba}{\begin{eqnarray}}
\newcommand{\ea}{\end{eqnarray}}

\newcommand{\ignore}[1]{}
\begin{document}
\begin{frontmatter}

\title {{Emergent complexity: what uphill analysis or downhill invention can not do}
\thanksref{label1}}\thanks[label1]{Working paper for a special issue of New Ideas in Psychology (2008). Work supported by NIH NINDS of USA (Grants 42660
and 35115) }
\author{Dante R. Chialvo}
\address{Department of Physiology, Feinberg School of Medicine , Northwestern University, 303 E. Chicago Avenue, Chicago, Illinois, 60611, USA.}

\begin{abstract}

In these notes we review emergent phenomena in complex systems, emphasizing ways to identify
potential underlying universal mechanisms that generates complexity. The discussion is centered
around the emergence of collective behavior in dynamical systems when they are poised near a
critical point of a phase transition, either by tuning or by self-organization. We then argue the
rationale for our proposal that the brain is naturally poised near criticality reviewing recent
results as well as the implications of this view of the functioning brain.
\end{abstract}


\end{frontmatter}

\section{Uphill analysis and downhill invention}
Decades ago, the paradigmatic Braitenberg's thought experiments in ``Vehicles, experiments in
synthetic psychology'' \cite{vehicles} already showed that ``vehicles'' with even rudimentary
internal structure can behave in surprisingly complex ways. He was able to demonstrate that by
wiring the vehicle few motors and sensors in different ways it can generate behaviors we might call
hate, aggression, love, foresight, or even optimism. In his own words:
\begin {quote}``It is pleasurable and easy to
create little machines that do certain tricks. It is also quite easy to observe the full repertoire
of behavior of these machines -- even it it goes beyond what we had originally planned, as it often
does. But it is much more difficult to start from the outside and try to guess internal structure
just form the observation of the data. [...] Analysis is more difficult than invention in the sense
in which, generally, induction takes more time to perform than deduction: in induction one has to
search for the way, whereas in deduction one follows a straightforward path. A psychological
consequence of this is the following: when we analyze a mechanism, we tend to overestimate its
complexity.''
\end{quote}
 Braitenberg was successful conveying the idea of a ``law of uphill analysis and
downhill invention'' capturing the  difficulty of guessing the internal structure from the
observation of behavior compared with the easiness of building artifacts exhibiting the behavior.

Recent work on complex systems allows for rather similar, albeit far reaching, suggestions. It is
now understood how robust behaviors can emerge in large nonlinear dynamical systems given some
minimal conditions. In some cases these behaviors -or  attractors, in the jargon- are universal
(e.g, can be equally seen in disparate domains) not only qualitatively but also in their
quantitative expressions. The intention of these notes is to link these results with the problem of
understanding brain function. It shares Braitenberg's motivation, without restricting our view of
the brain to low dimensional systems (i.e., two motors, two sensors, and a few ``neurons'').
Besides its realism, the high dimensionality assumption is preventing us from ``inventing'' any
structure, which then must emerge only from self-organizing dynamics. The emphasis is to see what
kind of behavior can these dynamical systems generate, under which constraints, etc. Much in the
same way that the vehicles exhibited love, hate and so on, we investigate the generic properties of
this scenario identifying those relevant to the problem of brain function.

The paper is organized as follows: Section 2 illustrates the idea of emergence in complex
systems,by describing previous work on the dynamics of social insects. Section 3 discusses what is
considered as complex followed by an account of the main theoretical efforts explaining natural
complexity as a result of being poised at a state of affairs close to criticality. Section 4
discusses how to get around Braitenberg's law, and which aspects of brain dynamics might benefit
from this viewpoint. The final comments are dedicated to review the arguments and recent results as
well as the implications of this view of the functioning brain.

\section{Emergence}
Emergence refers to the unexpected spatiotemporal patterns exhibited by complex systems. As
discussed at length elsewhere \cite{bakbook,maya,chialvo2004}, complex systems are usually
\emph{large} conglomerate  of \emph{interacting} individuals, each one exhibiting some sort of
\emph{nonlinear} dynamics. In addition, there is usually energy being pumped into the system, thus
some sort of driving is present. The three \emph{emphasized} components are necessary (although not
sufficient) conditions for a system to exhibit at some point emergent behavior. Suppose that the
elements are humans, who are driven by food, sun light and other energy sources, they can form
families and communities. At a certain point in their evolution, they give themselves some
political structure, with, say, chiefs and counsellors. Whatever the type of structure that emerge
it is unlikely to appear if one of the above emphasized elements is absent. It is well established
that a small number of isolated linear elements is not going to produce much unexpected behavior
(mathematically, this is the case in which everything can be formally anticipated).
\begin{figure}[htbp]
\vspace{10mm}
 \centering{\includegraphics[height=100 mm,angle=-90]{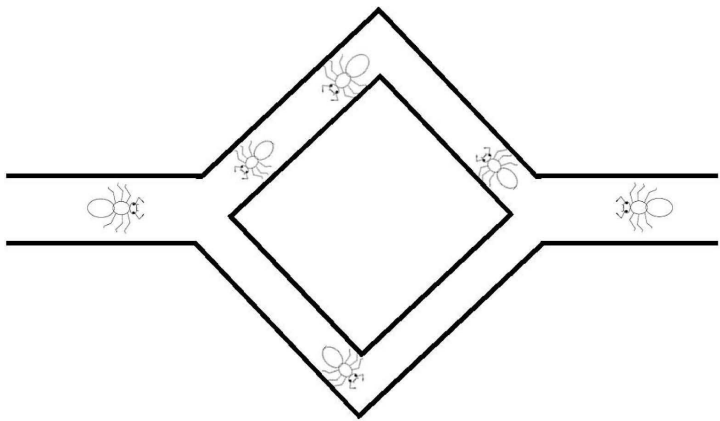}}
\caption{In the binary bridge experiments ants are exposed to bridges connecting two or more areas.
Eventually they discover and cross the bridges.}
\end{figure}
We will consider the dynamic of swarms, as a specific example with the three necessary components
which will help make the point clear. Social insects represent the paradigm of cooperative dynamics
\cite{wilson71}. In the case of foraging ants they interact to organize trails connecting the nest
with the food sources, forming structures with sizes that usually are several order of magnitude
larger than any of the individual's temporal or spacial scales. Therefore, the relevant problem
here is to understand the microscopic mechanism by which relatively unsophisticated ants, can build
and maintain these very large macroscopic structures.\footnote{We touch here only the surface of
this problem, the interested reader will find the full account in
\cite{rauch,chialvomillonas,millonas1,millonas2,millonas3} as well as the most recent extension of
this work to image processing \cite{ramos1} as well as to some fascinating non human art
\cite{ramos2}.}

To study this problem, Millonas introduced \cite{millonas1} a spatially extended model of what he
termed ``protoswarm''. In the swarm, there are two variables of interest, the organisms (in large
number) and a field, representing the spacial concentration of a scent (such as pheromone). As in
real ants, the model's organisms are influenced in their actions by the scent field and in turn
they are able to modify by depositing a small amount of scent in each step. The scent is slowly
evaporating as well. The organisms only interact trough the scent's field. The model is inspired in
the behavior of real ants, in which they are exposed to bridges connecting two or more areas where
the ants move, feed, explore, etc. Eventually they will discover and cross one of the bridges. As
it is illustrated in Figure 1 they will come to a junctions where they have to choose again a new
branch, and continue moving. Since ants both lay and follow scent as they walk, the flow of ants on
the bridges typically changes as time passes. In the example illustrated in Figure 1, after a while
most of the traffic will eventually concentrate on one of the two branches. The collective switch
to one branch is the emergent behavior, something that can be understood intuitively on the basis
of the positive feedback between scent following, traffic, and scent laying.  Numerous mathematical
models and computer simulations were able to capture this behavior observed in the laboratory as
well \cite{beckers,deneubourg1,pasteels}.
%%%%%%%%%%%%%%%%%%%%%%%%%%%%%%%%%%%%%%%%%%%%%%%%%%%%%%%%%%%%%%%%%
\begin{figure}[htbp]
 \centering{\includegraphics[height=100mm]{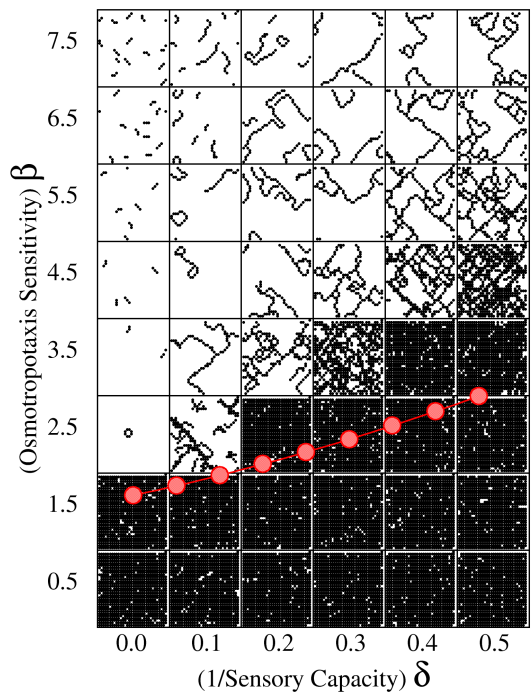}}
\caption{Emergent patterns exhibited by synthetic swarms with different values of gain ($\beta$)
and dynamic range ($\delta$) in the ants' sensory system. For each parameter $\beta$ and $\delta$
value indicated on the axis, the dots in each of the 48 squares is a snapshot of the ants'
positions at the last ten iterations after 5000 time steps starting from a random initial
distribution of agents. The circles joined by a line are the predicted values for the phase
transition. There are three distinct behaviors: one region (below the line) where behavior is
random, a second (above the line) where line of ordered traffic appears, and yet another with
clustered immobile ants (far above the line). (32x32 lattice, other parameters as in
\cite{rauch,chialvomillonas}.)}
\end{figure}
%%%%%%%%%%%%%%%%%%%%%%%%%%%%%%%%%%%%%%%%%%%%%%%%%%%%%%%%%%%%%%%%
The understanding gained with the bridge experiments can be extended to more sophisticated
settings, such as freely exploring organisms. The insight comes from a rather clever way that
Millonas's model discretized space. For descriptive purposes, the model can be considered as a
network constructed by connecting each point of a square lattice to its eight nearest neighbors
with the bridges of Figure 1. Thus, at each step (literally) each ant makes a decision to choose
one of eight bridges; and deposits a fixed amount of pheromone as it walks, that is all. The
decision is based on the scent amount at each of the eight locations. The ants' sensory apparatus
embedded in a physiological response function, was modelled following biological realism, having
two parameters, one which could be considered analogous to gain and the other  the inverse of
sensory capacity (or dynamic range). The plot in Figure 2 condenses the results from  the numerical
simulations with different values for the physiological response function. At each combination of
the explored $\beta$ and $\delta$ values there is a square plot which depicts the locations of each
ant at the last ten steps of the simulation. It can be seen that ants converge to different
behaviors depending on the parameter' values. For values of both small gain and small dynamic range
ants execute a random path, resulting in the plots fully covered, as in the right bottom corner.
That makes sense, because of the low sensitivity ants are just making random choices at each
juncture. For large enough gain (top left corner), ants senses saturate resulting in clusters of
immobile ants in the same attracting spot. It is in between these two states, one disordered and
the other frozen, that the swarm can organize and maintain large structures of traffic flow as
those seen in nature.
%%%%%%%%%%%%%%%%%%%%%%%%%%%%%%%%%%%%%%%%%%%%%%%%%%%%%%%%%%%%%%%%%
\begin{figure}[htbp]
 \centering{\includegraphics[height=100mm]{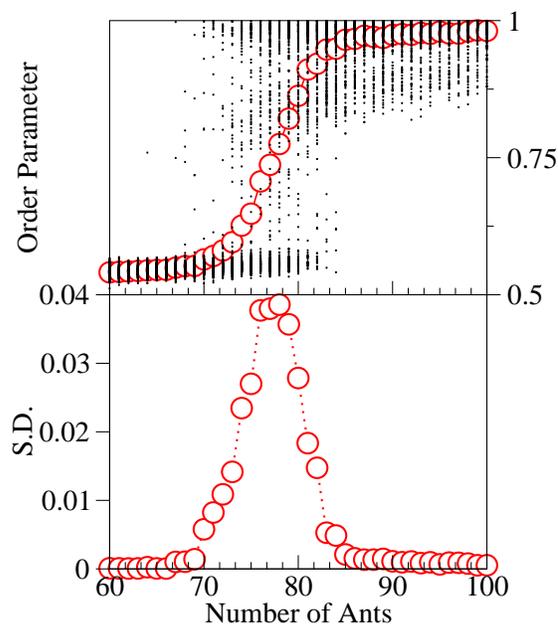}}
\caption{Collective swarm behavior near the phase transition. Collective order (top) increases as a
function of increasing density of ants  and fluctuations magnitude (bottom) diverges at the
critical point. (Order parameter values from individual runs are plotted using dots and circles for
the average. S.D: Standard deviation of the data in the top panel.)}
\end{figure}
%%%%%%%%%%%%%%%%%%%%%%%%%%%%%%%%%%%%%%%%%%%%%%%%%%%%%%%%%%%%%%%%
The stability analysis in the theory in \cite{millonas1} makes straightforward to understand the
transition between the disordered walks and the complex structures of trails (line and circles in
Fig. 2). Intuition already suggests that the model's $\beta$,$\delta$ values at which the
order-disorder transition happens must depend on the number of ants able to reinforce the scent
field. In Figure 3 results from several hundreds runs with fixed parameters ($\beta=3.5,
\delta=0.1$) and increasing density of ants are shown. The degree of collective order was evaluated
by the proportion of ants ``walking'' on lattice points having above average scent concentration
(see further details in \cite{rauch,chialvomillonas}). The top panel shows a plot of the results
where the dots indicate the outcome of each individual run (with different initial conditions) and
circles the average of all runs. For low ants' density the expected random behavior is observed,
with equal likelihood for ants to be in or out of a high scent field. For increasing number of ants
the swarm starts to order to reach the point in which the majority of the ants are walking on a
field with scent concentration larger than the average. Of note is the generic increase of the
amplitude of the fluctuations right at the transition, as it is shown in the bottom panel of Figure
3.

These results were the first to show the simplest (local, memoryless, homogeneous and isotropic)
model which leads to trail forming, where the formation of trails and networks of ant traffic is
not imposed by any special boundary conditions, lattice topology, or additional behavioral rules.
The required behavioral elements are stochastic, nonlinear response of an ant to the scent, and a
directional bias. There are other relevant properties, discussed in detail elsewhere
\cite{rauch,chialvomillonas}, that arise \emph{only} at the region of ordered line of traffic,
including the ability to reconstitute trails and amplify weak traces of scent. The main conclusion
to be drawn from the work commented here is that scent-following is sufficient to produce the
emergence of complex patterns of organized flow of social insect traffic all by itself (and not
just sufficient to allow for trail following behavior). Lets recall now the three aspects
emphasized at the beginning of this section as crucial for emergence. As shown in Fig. 3 a certain
minimum density of ants is needed since trails need to be reinforced, the response to the scent
need to be nonlinear, and ants must interact (through the field). These are all essential
ingredients for the emergence of these complex patterns of swarming behavior.

\section{What is complex?}
We have seen in the previous section that complex structures of trails emerge near a phase
transition. But, what it means for something to be complex? Except for man-made objects, complexity
\emph{is} what we see usually all around us. Instead of \emph{ad hoc} definitions, it is much more
telling to discuss how nature manages to build the complexity we are all embedded in. As the
opening sentence of Per Bak \cite{bakbook} book reminds us:
\begin{quote}\emph{"How can
the universe start with a few types of elementary particles at the big bang, and end up with life,
history, economics, and literature. The question is \textbf{screaming out to be answered} but it is
seldom even asked. Why did the big bang not form a simple gas of particles or condense into one big
crystal? We see complex phenomena around us so often that we take for granted without looking for
further explanation. In fact, until recently very little scientific effort was devoted to
understanding why nature is complex?"} \end{quote} The additional appeal of \emph{explaining}
complexity is that entails \emph{figuring out} mechanisms, nothing essentially different from the
long prolific tradition of physics in other more simple systems. That work provided us with all the
laws needed to understand a large body of problems. But that is history, the problem scientists are
faced with now is to write down the equations able to produce the surrounding complexity one sees.
Clouds, epidemics, rainy as well as drought seasons, flowers, storms, earthquakes, economics,
diversification and extinction of biological species, galaxies, societies, wars and peace are a few
examples of complex systems we need  laws for. Some, will argue as nonsensical to even treat
economics within the realm of physics, others that it is perfectly possible, and furthermore that
it won't be necessary to conceive as many laws as objects of interest. After all, they consider,
Newton's laws are unique; there is not one written for falling apples, another one for falling
airplanes, etc. Work in recent years shows that there are large classes of complex systems obeying
the same universal laws, making simpler the task of figuring out how they work. Nowadays, this
concept is more than intuition and there are concrete examples in which this universality have
already been demonstrated \cite{bakbook,Sole,Buchanan}.

\section{Emergent complexity is always critical}
Look around and observe what you see. It is fair to say that, for the most part, the objects and
phenomenology we do understand have a common feature: their regularity and uniformity. In contrast,
nature is by far \emph{non homogeneous, or non uniform}. This observation can not be brushed off
for trivial or exceptional, it is the rule and as Bak argued \cite{bakbook} a question screaming
out to be answered. Attempts to explain and generate this kind of non uniformity included many
mathematical models and recipes, but few succeeding in creating complexity without embedding the
equations with it. The point being that including the complexity in the model will only be a
simulation of the real system. That is not what we are after. We want rather to understand
complexity which implies to discover the conditions in which something complex emerges from the
interaction of the constituting non-complex elements.
%%%%%%%%%%%%%%%%%%%%%%%%%%%%%%%%%%%%%%%%%%%%%%%%%%%%%%%%%%
\begin{figure}
\includegraphics[height=140mm]{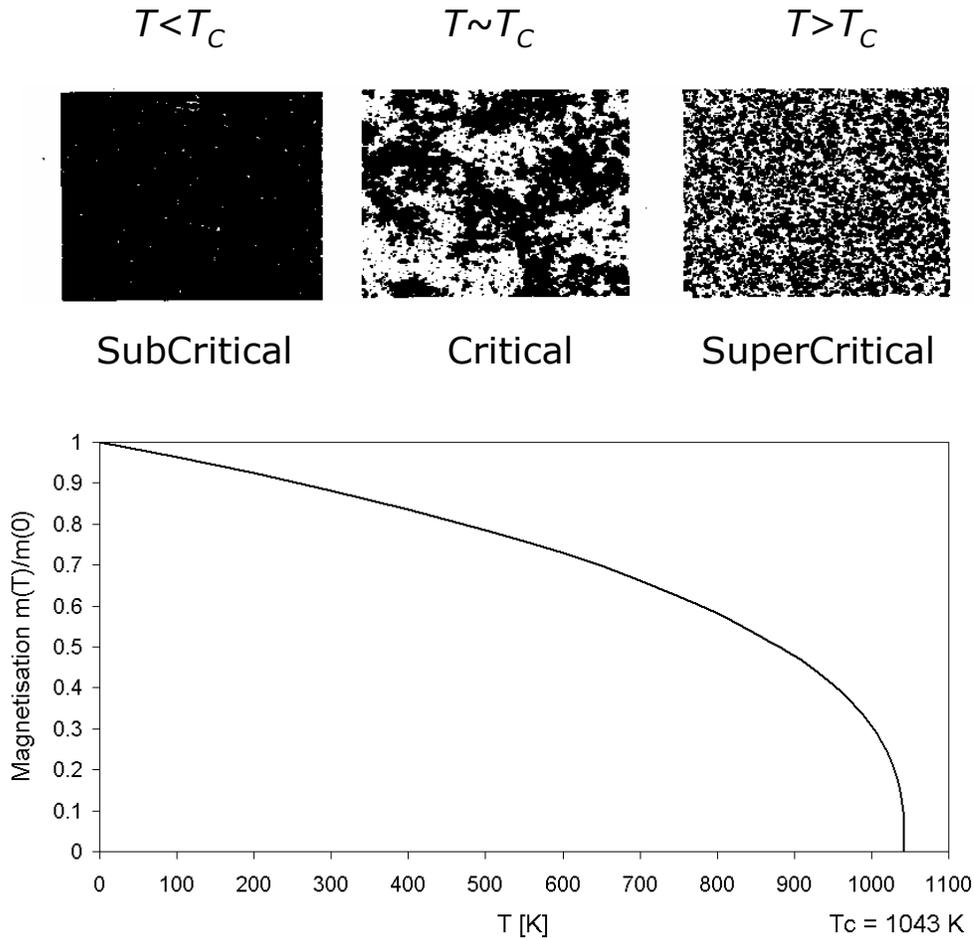}
\caption{Complex is critical: Example of a ferromagnetic-paramagnetic phase transition. Bottom:
Temperature dependence of magnetization m(T) for Fe. Top three panels are snapshots of the spins
configuration at one moment in time for three temperatures: subcritical, critical and supercritical
from numerical simulations of the Ising model (d=2). Notice that both sub- and supercritical
conditions result in homogeneous states, while at the critical temperature the system exhibits
highly inhomogenous correlated domains. This is typical for any system exhibiting a second order
phase transition.}
\end{figure}
%%%%%%%%%%%%%%%%%%%%%%%%%%%%%%%%%%%%%%%%%%%%%%%%%%%%%%%%%%%
Initial inspiration about the possible origins of natural complexity was provided by work in the
field of phase transitions and critical phenomena. To expand on these ideas lets recall the
scenario of ferromagnetic-paramagnetic phase transition illustrated in Figure 4.  A material is
said to be ferromagnetic if it displays a spontaneous magnetization in absence of any external
magnetic field. If, for instance, we heat up an iron magnet the magnetization gets smaller and
finally reaches zero, following the curve depicted in the figure. If one could see the orientations
of the individual spins, will realize that at low temperature the system is very ordered with many
very large domains of equally oriented spins, a state that is practically invariant in time. On the
other extreme, at very high temperatures, spins' orientation changes constantly, they are
correlated at only very short distances and as a consequence the mean magnetization. which
expresses the collective behavior, vanishes; hence, the iron is not a magnet anymore. In between
these two homogeneous states, at what is called the critical temperature $T_{c}$, the system
exhibits very peculiar fluctuations both in time and space. For example, the magnetization temporal
fluctuations are known to be scale invariant. Similarly, the spatial distribution of spins'
clusters show long range (power law) correlations. At the critical point, these large dynamic
structures emerge, even though there are only \emph{short-range} interactions between the systems'
elements. Thus, at the critical temperature, the system exhibits a greatly correlated (up to the
size of the system) state which at the same time is able to fluctuate wildly in time at all scales.

One would agree that, at least qualitatively, the spatiotemporal patterns observed at the critical
point resemble natural objects (say clouds or coastlines), an impression that could be reinforced
by further quantitative analysis of their self similarity both in time and space. This is also
consistent with the results in section 2 confirming that complex patterns of ants' traffic emerge
only near the critical line. However, given the fact that these properties are present \emph{only
at} the critical point, criticality can not be suggested as a source of complexity, unless a robust
mechanism to tune nature towards criticality is provided.

That was precisely Per Bak's contribution \cite{bakbook} with the theory of Self Organized
Criticality (SOC) \cite{bakbook}: to identify the probable mechanisms by which large dynamical
systems can drive themselves towards a critical point. The theory's -already two decades old- main
claim is that large non linear dynamical systems can have an intrinsic tendency to evolve
spontaneously toward a critical state characterized by spatial and temporal self-similarity, of the
kind seen to abound in nature. The crucial assumption in all SOC models is the interaction of many
nonlinear elements and the slow pumping of energy. In the original formulation, the basic idea
behind SOC was framed in a toy model, e.g., a \emph{thought} experiment, using a mathematical model
of a sand pile, inspired in the dynamics of avalanches seen in real piles of sand. To visualize the
concept, think of building a real sand pile. By dropping sand slowly, grain by grain on a surface,
eventually one reaches a situation in which the slope of the pile is not growing anymore. At this
point the pile is critical, i.e. it has a slope at which any further addition of sand will produce
sliding of sand (avalanches). This sliding can at times be very small or at some others can cover
the entire size of the system. One quickly realizes, that each avalanche mobilizes only enough sand
to decrease the local slope below the critical value. In that way the system \emph{tunes itself} to
a point by dissipating just enough energy (in the avalanches) to become barely subcritical, while
the slow but continuous pumping of energy assures the return towards criticality. This self
organized critical state is characterized by scale-invariant distributions for the size and the
lifetime of the avalanches.

Considering the problem at hand, one would imagine that the SOC model needs to be very elaborate,
but that it is not the case. The 2D version of the model runs on a square grid. Each site is
identified by a position $x$ and $y$ and its state solely characterized by a local ``slope''
$z(x,y)$. Starting with a flat surface $z(x,y) = 0$, at each step one chooses an arbitrary location
and drops a grain of sand, causing $z(x,y)=z(x,y)+1$. Anytime that $z$, at any location, is larger
than a threshold of 4, the $z$ value at that location is decreased by 4 and the four neighbors
$z'$s increased by 1, these are the avalanches. That is all!.

It is important to reflect for a minute on the properties of the self-organized critical state.
First of all, the behavior one observes in the sand pile experiment is highly non-uniform; that is,
most of the times nothing happens, infrequently big avalanches occur, but also everything in
between is possible with a probability inverse to the slides sizes.  In addition, the
non-uniformity happens for the same input, a single grain of sand. However unpredictable this could
sound it is not random behavior, since it is observed in a completely deterministic model. The
reason for the seemingly unpredictable dynamics lies in the high dimensionality of the system. When
we add a single grain, the future of the pile is written in the precise state of each of the other
millions of grains. This -only apparent- unpredictability is different from the dynamics of
deterministic chaos, which is also erratic behavior but produced by \emph{low dimensional}
deterministic systems. Among other contrasting properties, one should note that sand piles (and SOC
in general) have a long memory, in the sense that the pile is shaped by the previous history of
events (avalanches) while in deterministic chaos the system forgets exponentially its past. The
long memory have a spatial counterpart discussed in the context of magnetization above. Of note is
that the critical state is the most \emph{unstable} and at the same time the most \emph{robust}
attractor for the dynamics. The most unstable, because a single grain of sand have the potential of
producing an avalanche of the entire sand pile. The most robust, because the system will eventually
come back to the same unstable state. The simultaneous presence of both instability  and robustness
properties is an important peculiarity of SOC systems, not seen in any other dynamical system. It
is clear that the system is self-organized, and it is hard to identify where in the model (or in
the real pile) is predicted the emergence of these peculiar properties.

For the last two decades SOC has been studied well beyond the initial sandpile metaphor, in models
and experiments, in a wide diversity of disciplines from biology to economics, from astrophysics to
linguistic, shedding light over the mechanisms generating complexity (for recent reviews see
\cite{maya,turcotte}).  Thanks to this work, it is now well established that a large class of
driven nonlinear dynamical systems can evolve to a critical state in which the above discussed non
uniformity will spontaneously appear. The only necessary condiments are: 1) a large number of
degrees of freedom, 2) each one must be non linear and 3) energy needs to be pumped to the system
relatively slowly and will be dissipated as fractals in space and time. The common theme across
these models and systems is the emergence of complexity at the critical state.\footnote{Note that
similar observation was made in the swarm model of section 3, where complex structures of traffic
form \emph{only} at the vicinity of the critical phase transition.}

\section{A way out of Braitenberg's law dead end}
Recall the introductory paragraph emphasizing the obstacle to explain behavior in terms of the
underlying \emph{collective}. In Braitenberg's words it is much easier to invent something that
replicates the behavior than to figure out the mechanisms from its observation. But Braitenberg's
downhill invention won't help much as soon as one chooses to explain some more sophisticated
cognitive behavior. At the same time, reductionist analysis is immediately precluded by the large
size and the nonlinearities of the system. Apparently it is a dead end. \emph{The way out suggested
here is to ask what kind of behavior large dynamical systems will universally exhibits.} In other
words the strategy is to look at the fundamental laws for the collective of neurons, in the hope of
finding a family of self-organized invents! We can certainly anticipate the most probable
criticism: all we will be describing is the mind as an epiphenomena of the brain.

We can start the  search guided by some relevant facts as source of inspiration. The brain have, as
a collective, some notoriously conflictive demands. On one side it needs to be ``integrated'' while
at the same time being able to remain ``segregated'',  as discussed extensively by Tononi and
colleagues
 \cite{Tononi98a,Tononi98,Tononi2004}. This is a non trivial constraint, nevertheless mastered by the brain as
it is illustrated with plenty of neurobiological phenomenology. Suffice to think in any conscious
experience to immediately realize that it always comprises a single undecomposable integrated state
\cite{Tononi98}. This integration is such that once a cognitive event is committed, there is a
refractory period (of about 150 msec.) in which nothing else can be thought of. At the same time
the large number of conscious states that can be accessed over a short time interval exemplify very
well the segregation property. As an analogy, the integration property we are referring to could be
also interpreted as the capacity to act (and react) on an all-or-nothing mode, similar to an action
potential or a travelling wave in a excitable system. The segregation property could be then
visualized as the capacity to evoke equal or different all-or-nothing events using different
elements of the system. This could be more than a metaphor.

While the study of this problem is getting increasing attention, the mechanisms by which this
remarkable scenario can exist in the realms of brain physiology is not being discussed as much as
it should. Recalling the properties of the dynamical scenario described in section 4, we propose
that the integration-segregation capacity of the conscious brain is related to it being poised at
the critical point of a phase transition. It is important to note that there is no other
conceivable dynamical scenario or robust attractor known to exhibit these two properties
simultaneously. Of course, any system could trivially achieve integration and long range
correlations in space by increasing link's strength among faraway sites, but these strong bonds
would prevent any segregated state. At the critical point these and others properties -equally
crucial for brain function- appear naturally. If the concept is correct, statistical physics could
help to move the current debate from phenomenology to understanding of the lower level brain
mechanisms of cognition.

The brains we see today are here precisely because they got an edge to survive, in that sense how
consistent is the view of the brain near a critical point with these Darwinian constraints? The
brains are critical because the world in which they have to survive is up to some degree critical
as well. In a sub-critical world everything would be simple and uniform (as in the left panel of
Figure 4) and there would be nothing to learn; a brain will be completely superfluous. On the other
extreme, in a supercritical world everything would be changing all the time (as in the right panel
of Figure 4); there, it would be impossible to learn. Brains are only necessary to navigate in a
complex, critical world, where even the very infrequent events have still a finite opportunity to
occur.\footnote{As the very big avalanches in the SOC model, discussed in section 4.} In other
words we need a brain \emph{because} the world is critical \cite{bakbook,maya,bak1,bak2,bak3}.
Furthermore, a brain not only have to remember, but also to forget and adapt. In a sub-critical
brain' memories would be frozen. In a supercritical brain, patterns change all the time so that no
long term memory would be possible. To be highly susceptible, the brain itself has to be in the
in-between critical state.

A number of features, known to be exhibited by thermodynamic systems at the critical point, should
be immediately observed in experiments, including:

\begin{enumerate}
\item At large scale:\\
 Cortical long range correlations in space and time.\\
 Large scale anti-correlated cortical states.
\item At smaller scale:\\
 "Neuronal avalanches", as the normal homeostatic state for most neocortical
 circuits.\\
``Cortical-quakes'' continuously shaping the large scale synaptic landscape providing ``stability''
to the cortex.
\item At behavioral level:\\
 All adaptive behavior should be ``bursty'' and apparently unstable, always at the ``edge of failing''.\\
 Life-long learning should be critical due to the effect of continuously ``raising the bar''.
\end{enumerate}
In addition one should be able to demonstrate that a brain behaving in a critical world performs
optimally at some critical point, thus confirming the intuition that the problem can be better
understood considering the environment in which brains evolved.

In the list above, the first item concerns the most elemental facts about critical phenomena:
despite the well known \emph{short range} connectivity of the cortical columns, \emph{long range}
structures appear and disappear continuously, analogous to the long ant's trails shown to form near
the critical point in section 2. The presence of inhibition as well as excitation together with
elementary stability constraints lead to conclude that the spatiotemporal cortical dynamics should
exhibit large scale anti-correlated spatial patterns as well, as reported recently in
\cite{Fox2006}. The features at smaller scales could have been anticipated from theoretical
considerations as well, but avalanches were first observed empirically in cortical cultures and
slices by Plenz and colleagues \cite{Plenz04}. According to a recent review \cite{PlenzTINS}
neuronal avalanches can be conceived as the atom of neuronal ensembles.

According to recent work, our senses seems to be operating at a critical point as well. To move
around, to escape from predators, to choose a mate or to find food, the sensory apparatus is
crucial for any animal survival. But it seems that senses are also critical in the thermodynamic
sense of the world. Consider first the fact that the density distribution of the various forms of
energy around us is clearly inhomogeneous, at any level of biological reality \footnote{If one
comfortably accept these facts then is left to believe that the world as a whole is critical.},
from the sound loudness any animal have to adapt to, the amount of rain a vegetal have to take
advantage. From the extreme darkness of a deep cave to the brightest flash of light there are
several order of magnitude changes, nevertheless our sensory apparatus is able to inform the brain
of such changes. It is well known that isolated neurons are unable to do that because of their
limited dynamic range, which spans only a single order of magnitude. This is the oldest unsolved
problem in the field of psychophysics, tackled very recently by Kinouchi and Copelli
\cite{kinouchi} by showing that the dynamics emerging from the \emph{interaction of coupled
excitable elements} is the key to solve the problem. Their results show that a network of excitable
elements set precisely at the edge of a phase transition - or, at criticality - can be both,
extremely sensitive to small perturbations and still able to detect large inputs without
saturation. This is generic for any network regardless of the neurons' individual sophistication.
The key aspect in the model is a local parameter that controls the amplification of any initial
firing activity. Whenever the average amplification is very small activity dies out; the model is
subcritical and not sensitive to small inputs. On the other hand, choosing an amplification very
large one sets up the conditions for a supercritical reaction in which for any - even very small -
inputs the entire network fires. It is only in between these two extremes that the networks have
the largest dynamic range. Thus, amplification around unity, i.e., at criticality, seems to be the
optimum condition for detecting large energy changes as an animal encounters in the real world
\cite{chialvo2006}. It is only in a critical world that energy is dissipated as a fractal in space
and time with the characteristic highly inhomogeneous fluctuations. Since the world around us
appears to be critical, it seems that we, as evolving organisms embedded in it, have no better
choice than to be the same.

At the next level the suggestion is  that human (and animal \cite{boyer}) behavior itself should
show indications of criticality. Learning obviously  must be considered candidate to be critical as
well, if one realizes that for teaching any skill one chooses increasing challenge levels, which
are easy enough to engage the pupils but difficult enough not to bores them. This ``raising the
bar'' effect continues trough out life, pushing the learner continuously to the edge of failure! It
would be interesting to measure some order parameter for sport performance to see if it shows some
of these features for the most efficient teaching strategies.

 %%%%%%%%%%%%%%%%%%%%%%%%%%%%%%%%%%%%%%%%%%%%%%%%%%%%%%%%%%%%%%%%%%
\begin{figure}
\centering{\includegraphics[height=150mm]{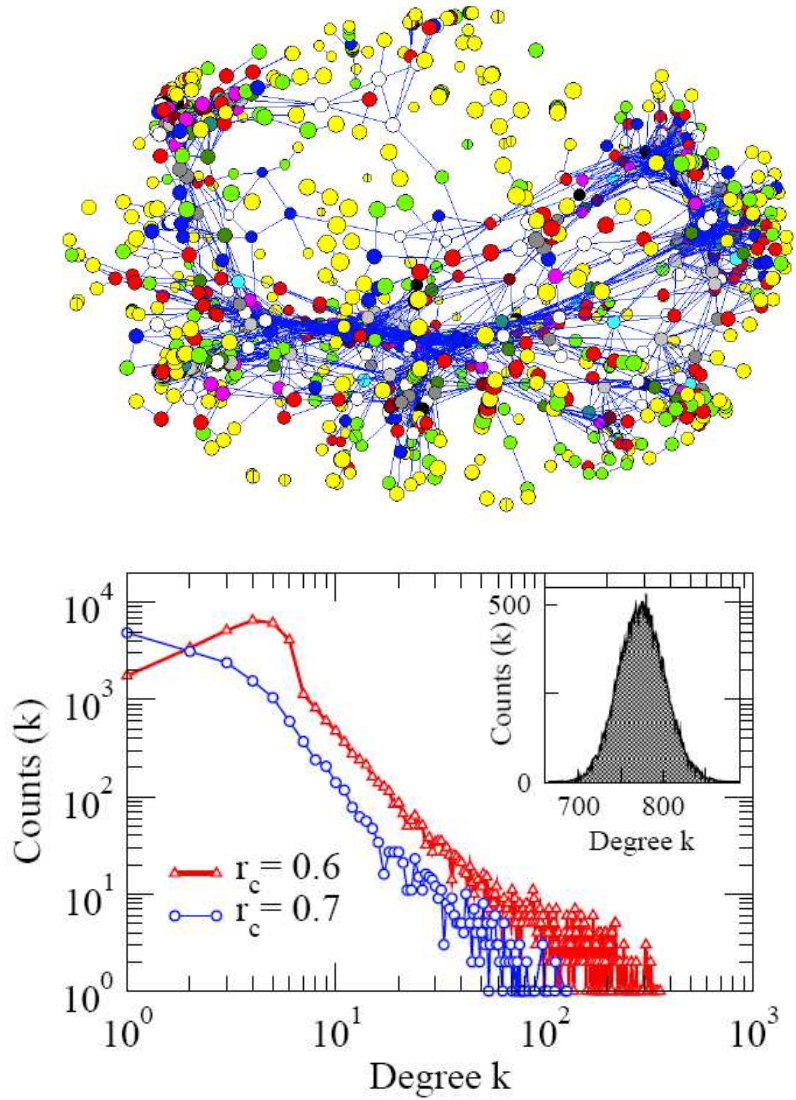}} \caption{A
typical brain network extracted from functional magnetic resonance
imaging. Top panel shows a pictorial representation of the
network. The bottom panel shows the degree distribution for two
correlation thresholds $r_c$. The inset depicts the degree
distribution for an equivalent randomly connected network. Data
re-plotted from \cite{Eguiluz}.}
\end{figure}
%%%%%%%%%%%%%%%%%%%%%%%%%%%%%%%%%%%%%%%%%%%%%%%%%%%%%%%%%%%%%%%%
\section{Functional brain networks are complex} Functional magnetic resonance imaging (fMRI)
allows to monitor non invasively spatio-temporal brain activity  under various cognitive
conditions. Recent work using this imaging technique demonstrated complex functional networks of
correlated dynamics responding to the traffic between regions, during behavior or even at rest (see
methods in \cite{Eguiluz}). The data is analyzed in the context of the current understanding of
complex networks (for a review see \cite{Sporns2004}). During any given task the networks are
constructed first by calculating linear correlations between the time series of brain activity in
each of $36\times64\times64$ brain sites. After that, links are said to exist between those brain
sites whose temporal evolutions are correlated beyond a pre-established value $r_c$.

Figure ~5, shows a typical brain functional network extracted with this technique. The top panel
illustrates the interconnected network's nodes and the bottom panel shows the statistics of the
number of links (i.e., the degree) per node.  Note the high non-uniformity in the interaction
between nodes. There is a few very well connected nodes in one extreme and a large number of nodes
with a single connection. The typical degree distribution approaches a power law distribution with
an exponent around 2. Other measures revealed that the number of links as a function of -physical-
distance between brain sites also decays as a power law, something already confirmed by others
\cite{Salvador} using different techniques. Two statistical properties of these networks, path
length and clustering were computed as well. The path length ($L$) between two brain locations is
the minimum number of links necessary to connect both sites. Clustering ($C$) is the fraction of
connections between the topological neighbors of a site with respect to the maximum possible.
Measurements of $L$ and $C$ were also made in a randomized version of the brain network. $L$
remained relatively constant in both cases while $C$ in the random case resulted much smaller,
implying that brain networks are ``small world'' nets, a property with several implications in
terms of cortical connectivity, as discussed further in \cite{Sporns2004b,Sporns2004}. In summary,
the work in \cite{Eguiluz} showed that functional brain networks exhibit highly inhomogeneous scale
free functional connectivity with small world properties. Although these results admit a few other
interpretations, the long range correlations demonstrated in these experiments are consistent with
the picture of the brain operating near a critical point. Recent work from Achard and colleagues
\cite{achard} sheds light over these aspects. They analyze fMRI time series from 90 cortical and
subcortical regions acquired from healthy volunteers in the resting state. Using similar methods
they find similar small-world topology of sparse connections in the graphs of brain functional
networks, most salient in the low-frequency range. They report an exponentially truncated power law
for the degree distribution, and when the network was tested for damage, (by deleting nodes and
recalculating topological measures) they found that it was ``more resilient to targeted attack on
its hubs than a comparable scale-free network, but about equally resilient to random error.'' Thus,
they have not only confirmed and extended the initial observations of small world connectivity but
also characterized further higher order topological features of the extracted networks. Most
recently, Basset and colleagues \cite{basset2006a,basset2006b} analyzed the topology and
synchronizability of frequency-specific brain functional networks using wavelet decomposition of
magnetoencephalographic time series recordings concluding that ``human brain functional networks
demonstrate a fractal small-world architecture that supports critical dynamics and task-related
spatial reconfiguration while preserving global topological parameters''. Stam and colleagues has
been actively looking at functional brain networks defined from electroencephalogram. In the most
recent work \cite{stam} they reported ``loss of small-world network characteristics'' in
Alzheimer's disease patients. As more detailed knowledge of the properties of healthy and abnormal
brain functional networks is achieved, the need to integrate this data in a cohesive picture grows,
as discussed recently by Sporns and colleagues \cite{Sporns2006}.

\section{Outlook}
In these notes, we have attempted to -provocatively- inject some connections between ubiquitous
dynamics of complex systems and brain cognition and behavior. In that spirit we also recognize that
the most useful connections are also the most tenuous and hard to perceive at first. It is also
probable that it is yet premature times for brain theory. We had already pictured brain theory at a
stage comparable to physics in ``pre-thermodynamic'' times \cite{Chialvo2007}. Imagine yourself in
days previous to the notion of temperature. Similarities between scalding water and ice will be
supported by their similar ``burning'' (to the touch) properties, when hot or cold were only
subjective quantities. Of course, the notion of pressure and temperature together with the
identification of phases changed everything. There is plenty of room for optimism that brain theory
will eventually undergo similar transformation as Werner's recent perspective \cite{Werner2006}
seems to convey.

\vfill\eject

\end{document}